\documentclass[pdflatex,sn-mathphys-num]{sn-jnl}

\usepackage{graphicx}%
\usepackage{multirow}%
\usepackage{amsmath,amssymb,amsfonts}%
\usepackage{amsthm}%
\usepackage{mathrsfs}%
\usepackage[title]{appendix}%
\usepackage{xcolor}%
\usepackage{textcomp}%
\usepackage{manyfoot}%
\usepackage{booktabs}%
\usepackage{algorithm}%
\usepackage{algorithmicx}%
\usepackage{algpseudocode}%
\usepackage[normalem]{ulem}%
\usepackage{listings}%
\usepackage{caption}
\usepackage{subcaption}

\theoremstyle{thmstyleone}%
\theoremstyle{thmstyletwo}%
\newcommand{\pD}[2]{\frac{\partial #1}{\partial #2}}
\newcommand{\Er}{e_\mathrm{r}}

\theoremstyle{thmstylethree}%

\begin{document}

\title[Article Title]{High Energy Density Radiative Transfer in the Diffusion Regime with Fourier Neural Operators}


\author*[1]{\fnm{Joseph} \sur{Farmer}}\email{jfarmer4@nd.edu}

\author[1]{\fnm{Ethan} \sur{Smith}}\email{esmith45@nd.edu}
\equalcont{These authors contributed equally to this work.}

\author[1]{\fnm{William} \sur{Bennett}}\email{wbennet2@nd.edu}
\equalcont{These authors contributed equally to this work.}

\author[1]{\fnm{Ryan} \sur{McClarren}}\email{rmcclarr@nd.edu}
\equalcont{These authors contributed equally to this work.}

\affil*[1]{\orgdiv{Aerospace and Mechanical Engineering}, \orgname{University of Notre Dame}, \orgaddress{\city{Notre Dame}, \state{Indiana}, \country{USA}}}


\abstract{Radiative heat transfer is a fundamental process in high energy density physics and inertial fusion. Accurately predicting the behavior of Marshak waves across a wide range of material properties and drive conditions is crucial for design and analysis of these systems. Conventional numerical solvers and analytical approximations often face challenges in terms of accuracy and computational efficiency.  In this work, we propose a novel approach to model Marshak waves using Fourier Neural Operators (FNO). We develop two FNO-based models: (1) a base model that learns the mapping between the drive condition and material properties to a solution approximation based on the widely used analytic model by Hammer \& Rosen (2003), and (2) a model that corrects the inaccuracies of the analytic approximation by learning the mapping to a more accurate numerical solution. Our results demonstrate the strong generalization capabilities of the FNOs and show significant improvements in prediction accuracy compared to the base analytic model.}




\maketitle

\section{Introduction}\label{sec1}

Marshak waves, a common type of 
driven supersonic radiative heat waves,
play a key part in the physics of internal confinement fusion (ICF) \cite{lindl2004physics, rosen2005analytic, cohen2004tracer, lindl1995development}, astrophysics \cite{tranchant2022new, gonzalez20092d, rosen1996science} and other high energy density phenomena \cite{Drake}. In most cases, a full description of the radiative transfer process is not required. Therefore, approximations are in order. The diffusion approximation is one of these and is considered the simplest \cite{olson2000diffusion}. 

In some cases, analytic solutions to the radiation diffusion equation can be useful in understanding experiments \cite{hurricane2006bent, hammer2003consistent, heizler2016self, marshak, petschek1960penetration, lane2013new,krief2024self}. These analytic or semi-analytic models can be thought of as a reduced order approximation of the full system, which is itself a simplification. As examples, \cite{hurricane2006bent} reduces a two dimensional diffusion system via asymptotic expansion. The diffusion system is an approximation to higher order radiation transport equations. Marshak, the namesake of these waves, reduced a partial differential equation (PDE) into an ordinary differential equation (ODE) \cite{marshak, petschek1960penetration}.

Reduced order solutions have the benefit of simpler calculation, as solving an ODE is usually preferable to solving a PDE, and they can be interrogated to clarify physical relationships between parameters. However, coming to a semi-analytic or analytic solution often involves invoking simplifications which may debase the accuracy of the prediction.
Thus, the motive for this inquiry is to take a widely used and appreciated semi-analytic diffusion model, the Hammer and Rosen Marshak wave model (HR) \cite{hammer2003consistent}, and provide a correction to the model's limiting assumptions in a computationally efficient manner. 

 Classical numerical solvers such as finite difference, finite element, or finite volume methods discretize continuous equations into a finite set of algebraic equations \cite{renardy2006introduction, jones2009differential, sommerfeld1949partial, hughes2012finite, strikwerda2004finite, eymard2000finite}. These numerical solvers can be computationally expensive for high dimensional problems and for domains with complex geometries. In recent years, approaches that leverage ML have garnered support to alleviate these challenges \cite{wu2020data, navaneeth2022koopman, goswami2020transfer}. 

In particular, neural operators, a class of ML models, have emerged as a promising solution to these challenges. These operators learn mappings between infinite-dimensional function spaces, effectively approximating differential or integral operators that govern PDEs in a data driven manner \cite{li2020neural, lu2022comprehensive}. One of the key advantages of neural operators is that they only need to be trained once to learn a family of PDEs, and obtaining a solution for a new instance of a PDE parameter requires only a forward pass of the network.  Furthermore, neural operators are discretization-invariant as they share network parameters across discretizations, allowing for the transfer of solutions between meshes.   

The Fourier neural operator (FNO) \cite{li2020fourier} is a seminal neural operator that learns network parameters in Fourier space.  The FNO uses fast Fourier transform (FFT) for spectral decomposition of the input and computation of the convolution integral kernel in the Fourier space. This approach has shown promising results in learning the underlying physics of various PDEs including Burgers, Darcy, and Navier-Stokes equations. 

In this work, we propose to use FNO to learn the physics of Marshak waves for various input-output pairs. We develop two models: a base model which takes the physical parameters of the Marshak wave problem as input and outputs the time dependent wavefront position and temperature distribution as given by the HR model, and a hybrid approach which corrects the analytic HR solution to output the numerical solution to the full flux-limited diffusion equation.

The structure of this paper is as follows. The diffusion model for Marshak waves is introduced in Section \ref{sec2}. Hammer and Rosen's approximation is summarized in Section \ref{sec3}. The neural network that is employed to correct the HR model is discussed in Section \ref{sec4}. Finally, results and conclusions are offered in Sections \ref{sec5} and \ref{sec6}.

\section{Marshak wave problem}\label{sec2}
We study radiation diffusion in planar geometry, which assumes variation of the dependent variables only in a single direction, $x$. The evolutions of the radiation and material energy density are governed by \cite{bingjing1996benchmark},
\begin{equation} \label{eq:rad}
   \pD{\Er}{t} = \frac{\partial}{\partial x} \frac{c}{3\kappa(\rho, T)} \frac{\partial \Er}{\partial x}  + c  \kappa (a T^4 - \Er),
\end{equation}
\begin{equation} \label{eq:mat}
   \pD{e}{t} =  c  \kappa (e - a T^4 ) 
\end{equation}
where, $\Er$ is the energy density of the radiation and $e$ is the energy density of the material. $c$ is the speed of light, $\kappa$ is the opacity with units of inverse length, $a$ is the radiation constant, defined $a \equiv \frac{4\sigma}{c}$ where $\sigma$ is the Stefan-Boltzmann constant. $T$ is the material temperature and $\rho$ is the material density.

A Marshak boundary condition will specify the incoming radiation flux \cite{bingjing1996benchmark}, 

\begin{equation}\label{eq:BC}
    \Er(x=0, t) - \left(\frac{2}{3\kappa}\pD{\Er}{x}\right)\bigg{|}_{x=0} = \frac{4}{c}F_\mathrm{inc}.
\end{equation}
where $F_\mathrm{inc}$ is the incident flux on the surface at $x=0$. The material energy density is found via integration of the specific heat,
\begin{equation}
    e = \int_0^T\!dT' \:C_\mathrm{v}(T').
\end{equation}

Solutions to Eq.~\eqref{eq:rad} in the optically thick limit are recognizable by sharp drops in temperature near the wavefront and gradual temperature variation behind the front. This is because the radiation temperature and material temperature are in equilibrium behind the wavefront. Thus, is often valid to assume equilibrium between the radiation temperature and and material temperature, i.e. $\Er = a T^4$. This assumption simplifies Eqs.~\eqref{eq:rad} and \eqref{eq:mat} to a single equation for the material temperature, 
\begin{equation}\label{eq:diffusion_equilibrium}
    \pD{e}{t} = \frac{4}{3}\frac{\partial}{\partial x} \frac{1}{\kappa(\rho, T)}\left( \frac{\partial}{\partial x} \sigma T^4  \right)
\end{equation}
with the boundary condition at the surface, 
\begin{equation}
    T(x=0, t) = T_s(t).
\end{equation}
Furthermore, the equation of state is specified so that, 
\begin{equation}\label{eq:eos}
     e = fT^\beta\rho^{-\mu},
\end{equation}
This is the formulation given in \cite{hammer2003consistent}. The parameters $f,\beta,\mu$ are found by fitting experimental data, as in \cite{cohen2020key}. 


\section{Hammer and Rosen approximation}\label{sec3}

The Hammer and Rosen model for supersonic thermal radiation diffusion is a perturbative, semi-analytic, one dimensional solution to the diffusion equation under mild limiting assumptions. In particular, this model assumes planar geometry, power law representations for the opacity, $\frac{1}{K} = gT^\alpha\rho^{-\lambda}$, and material internal energy, $e = fT^\beta\rho^{-\mu}$, and a constant density. These assumptions transform Eq.~\eqref{eq:diffusion_equilibrium} into, 
\begin{equation}
    \rho \frac{\partial e}{\partial t} = \frac{4}{3} \pD{}{x} \left( \frac{1}{K\rho}   \pD{}{x} \sigma T^4\right ),
\end{equation}
where $\rho$ is the material density, $e$ is the internal energy, $\sigma$ is the Stefan-Boltzmann constant, and $T$ is the radiation temperature. The application of these assumptions and some simplification leads to the expression

\begin{equation}
    \frac{\partial T^\beta}{\partial t} = C\frac{\partial^2 }{\partial x^2}T^{4+\alpha}
\end{equation}
where our constants are collected into the term 
\begin{equation}
C = \frac{4}{4+\alpha}\frac{4}{3}\frac{1}{f}g\rho^{\mu-2-\lambda}
\end{equation}

This model predicts the position of the wave front as a function of time as the solution to an integral expression, then provides an explicit expression for the temperature profile in the material. The model can accommodate an arbitrary radiation temperature boundary condition. The Hammer and Rosen model gives the position of the wavefront, $x_\mathrm{f}$, as
\begin{equation}
    x_\mathrm{f}^2(t) = \frac{2+\epsilon}{1-\epsilon}CT_s^{-\beta}\int_0^t T_s^{4+\alpha} d\hat t
\end{equation}
where $T_s$ is the boundary temperature, $\epsilon = \frac{\beta}{4+\alpha}$ is a combination of terms from the power laws, and $x_\mathrm{f}$ is the heat front position as a function of time, $t$. With knowledge of the wavefront position a simple expression can be evaluated for the temperature profile:
\begin{equation}
\frac{T^{4+\alpha}}{T_s^{4+\alpha}}(x,t) = \left[\left(1-\frac{x}{x_\mathrm{f}}\right)\left(1+\frac{\epsilon}{2}\left(1-\frac{x_\mathrm{f}^2}{CH^{2-\epsilon}}\frac{dH}{dt}\right)\frac{x}{x_\mathrm{f}}\right)\right]^{1/(1-\epsilon)}.
\end{equation}
Here $H = T_s^{4+\alpha}$.
One hallmark of this approximate solution is that it is very inexpensive to evaluate. In practice, and when compared to computing a numerical solution, this method is effectively immediate. For this reason, it has proven to be particularly helpful for rapid iteration during the design process.

\section{Fourier neural operator model}\label{sec4}
We now turn to the consideration of producing a machine learning model to compute Marshak wave solutions. For this task we turn to the Fourier Neural Operator. In this section we use standard notation from the ML literature; regrettably, this overlaps with the standard notation for Marshak waves at times. 

\begin{figure}[!htbp]
    \centering
    \includegraphics[width=1.2\textwidth]{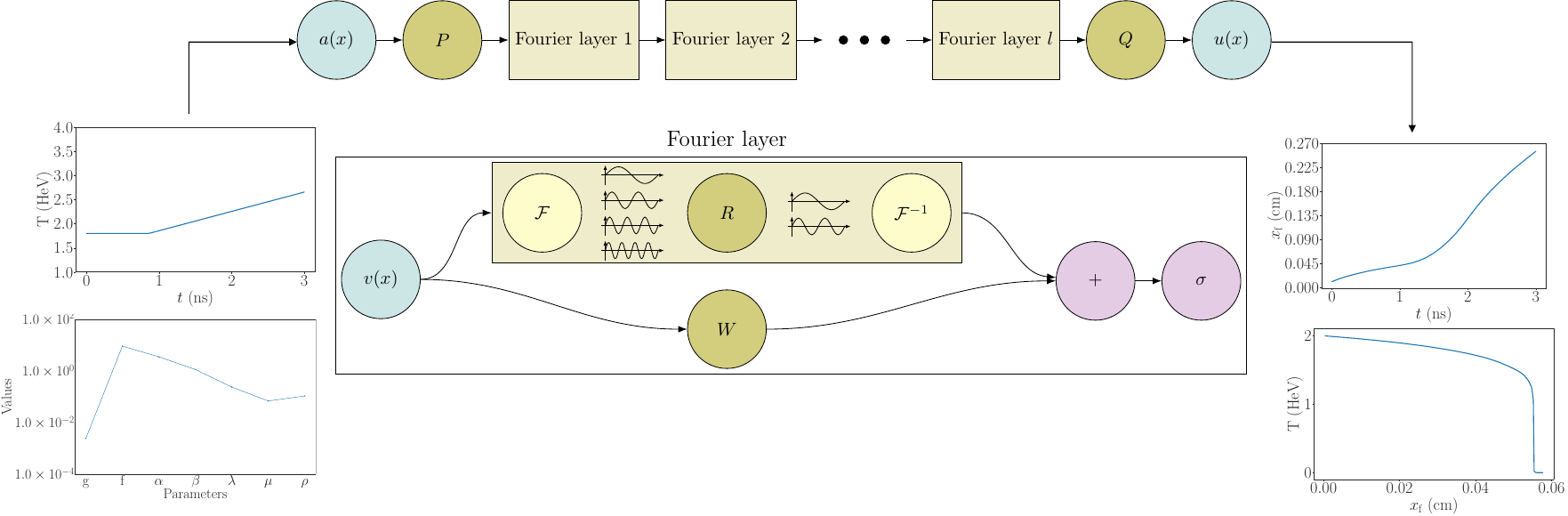}
    \caption{Fourier neural operator architecture for solving the Marshak wave problem.  The input function $a(x)$ is projected to a higher representation $v_0(x)$ by the projection layer $P$. This is then processed through $l$ iterations of Fourier layers. Each Fourier layer consists of a Fourier transform $\mathcal{F}$ that maps $v_i(x)$ to the Fourier domain, multiplication with the weight tensor $R$ and filtering of higher Fourier modes, and an inverse Fourier transform $\mathcal{F}^{-1}$ to return to the spatial domain.  The output is linearly transformed by $W$ and passed through a nonlinear activation function $\sigma$. This is added to the previous Fourier layer's output to produce the updated representation $v_{i+1}(x)$. After $l$ layers, the final representation $v_l(x)$ is mapped to the output solution $u(x)$. The boundary temperature drive (top left) and parameters (bottom left) represent the input functions and the front position (top right) and temperature distribution (bottom right) represent the output functions for the Marshak wave problem}
    \label{fig:FNO}
\end{figure}

The primary goal of an operator $\mathcal{G}$ is to establish a mapping between infinite-dimensional spaces from a finite collection of input-output pairs, denoted as $\mathcal{A} = A(\mathbb{R}^{d_a}) \subset \mathbb{R}^{d_a}$ and $\mathcal{U} = U(\mathbb{R}^{d_u}) \subset \mathbb{R}^{d_u}$, respectively. Following from \cite{tripura2022wavelet, li2020fourier}, consider a partial differential equation (PDE) which maps input function spaces to an output solution space. For a given domain $D \subset  \mathbb{R}^{d}$ with boundary $\partial D$, and $x \in D$, an operator would map source terms, $f(x, t) : D \to \mathbb{R}$, boundary conditions, $u(\partial D, t) : D \to \mathbb{R}$, and initial conditions $u(x, 0) : D \to \mathbb{R}$, to the solution space $u(x, t) : D \to \mathbb{R}$, where $t$ is time. In the present work, we aim to learn the nonlinear differential operator $\mathcal{G} : \mathcal{A} \to \mathcal{U}$ for various sets of input parameters $a \in \mathcal{A}$ in the Marshak wave problem. By constructing a parametric map $G: A \times \Theta \to U$, the optimal parameter $\theta \in \Theta$ can be approximated with data-driven methods to adjust $\theta$ such that $G(\cdot, \theta)$ approaches the target map $G$. Classical numerical solvers, be it finite elements, finite differences, or many modern data-driven and physics-informed neural networks attempt to learn the output function $u(x, t)$ which satisfies $\mathcal{G}$ for a single instance of input parameter $a$ and can be computationally prohibitive, especially when the solution for the PDE is required for many instances of the parameter. On the other hand, Fourier neural operators (FNO) have been developed to approximate $\mathcal{G}$ directly so that solutions to a family of PDEs are realized for different sets of $a$, thereby enhancing computational efficiency and practical utility.  

In general, input and output functions $a$ and $u$ are continuous, however, we assume to know only point-wise evaluations. To that end, the problem at hand can be described using the $n$-point discretization of $D$, $D_j=\left\{x_1, \ldots, x_n\right\} \subset D$ with observations of input-output pairs indexed by $j$ $\left\{a_j \in \mathbb{R}^{n \times d_a}, u_j \in \mathbb{R}^{n \times d_u}\right\}_{j=1}^N$, and $u_j = \mathcal{G}(a_j)$. The neural operator to learn the input-output mapping is an iterative architecture. First, the input $a(x,t)$ is transformed to a higher dimensional representation by $v_0(x) = P(a(x))$ where the transformation $P(a(x)) : \mathbb{R}^{d_a} \mapsto \mathbb{R}^{d_v}$. In this framework, a shallow fully connected network can achieve this desired transformation. Next a series of $l$ updates $v_i \mapsto v_{i+1}$ are performed  
\begin{equation}\label{eq:iter}
v_{i+1}(x) := \sigma \left(W v_i(x) + \left(\mathcal{K}(a; \phi)v_i\right)(x)\right), \quad \forall x \in D.
\end{equation}
with nonlinear activation function  $\sigma(\cdot) : \mathbb{R} \mapsto \mathbb{R}$ and a linear transformation $W : \mathbb{R}^{d_v} \mapsto \mathbb{R}^{d_v}$.  Each $v_i$ is a $d_v$-dimensional real vector in $\mathbb{R}^{d_v}$. For a vector input $x = [x_1,x_2, \ldots, x_{d_v}]^T \in  \mathbb{R}^{d_v}$, $\sigma(x)$ is applied element-wise, resulting in $[\sigma(x_1), \sigma(x_2), \ldots, \sigma(x_{d_v})]^T$. The integral kernel operator $\mathcal{K}: \mathcal{A} \times \theta \rightarrow \mathcal{L}(\mathcal{U}, \mathcal{U})$ is parameterized by $\phi \in \Theta_{\mathcal{K}}$
\begin{equation}\label{eq:integ}
\left(\mathcal{K}(a; \phi)v_i\right)(x) := \int_{D} \kappa_\phi(x, y, a(x), a(y); \phi) v_i(y) dy, \quad \forall x \in D.
\end{equation}
where $\kappa_\phi : \mathbb{R}^{2(d+d_a)} \to \mathbb{R}^{d_v \times d_v}$ is a neural network parameterized by $\phi \in \Theta_{\mathcal{K}}$. After all iterations, a transformation function $u(x)=Q\left(v_l(x)\right)$ moves $v_l(x)$ into the solution space $Q\left(v_l(x)\right): \mathbb{R}^{d_v} \mapsto \mathbb{R}^{d_u}$. This approach extends the idea of neural networks to operate on infinite-dimensional function spaces, enabling the learning of mappings between such spaces from finite data samples.  By leveraging neural operators, it becomes possible to approximate the nonlinear operators that govern the relationships between infinite-dimensional input and output function spaces, such as those arising in the context of partial differential equations. 

The FNO is a specific neural operator architecture designed for such nonlinear mappings. It replaces  the kernel integral operator in \label{eq:integ} by a Fourier convolution operator $\mathcal{F}^{-1}\left(\mathcal{F}\left(\kappa_\phi\right) \cdot \mathcal{F}\left(v_i\right)\right)(x)$, and applying the convolution theorem. The Fourier kernel integral operator becomes
$$
\left(\mathcal{K}(\phi) v_i\right)(x)=\mathcal{F}^{-1}\left(R_\phi \cdot\left(\mathcal{F} v_i\right)\right)(x), \quad \forall x \in D,
$$
where $\mathcal{F}$ is the Fourier transform of a function and $\mathcal{F}^{-1}$ is its inverse transform, $R_\phi$ is the Fourier transform of a periodic function $\kappa$ parameterized by $\phi \in \Theta_{\mathcal{K}}$. Given that $\kappa$ is periodic and can be represented by a Fourier series expansion, only discrete modes are considered $k \in \mathbb{Z}^d$. To create a finite dimensional representation, the Fourier series is truncated at a maximum number of modes $k_{\text{max}} = |\{ k \in \mathbb{Z}^d : |k_j| \leq k_{\text{max},j} \text{ for } j = 1, \ldots, d \}|$.

In a discretized domain $D$ with $n \in \mathbb{N}$ points, $ v_i \in \mathbb{R}^{n \times d_v }$ and $\mathcal{F}(v_i) \in \mathbb{C}^{n \times d_v}$ is obtained, here $\mathbb{C}$ represents the complex space. A convolution of $v_i$ with a function that has $k_{\text{max}}$ Fourier modes  gives $\mathcal{F}(v_i) \in \mathbb{C}^{k_{\text{max} \times d_v}}$. Then the multiplication with the weight tensor $R \in \mathbb{C}^{k_{\text{max} \times d_v \times d_v}}$ is  
\begin{equation}
\left(R \cdot\left(\mathcal{F} v_i\right)\right)_{k, l}=\sum_{j=1} R_{k, l, j}\left(\mathcal{F} v_i\right)_{k, j}, \quad k=1, \ldots, k_{\max }, \quad j=1, \ldots, d_v
\end{equation}
With uniform discretization and resolution $s_1 \times \cdots \times s_d=n$, Fast Fourier Transform (FFT) can replace $\mathcal{F}$.  For $f \in \mathbb{R}^{n \times d_v}, k=\left(k_1, \ldots, k_d\right) \in \mathbb{Z}_{s_1} \times \cdots \times \mathbb{Z}_{s_d}$, and $x=\left(x_1, \ldots, x_d\right) \in D$, the FFT $\hat{\mathcal{F}}$ and its inverse $\hat{\mathcal{F}}^{-1}$ are defined as

\begin{align}
& (\hat{\mathcal{F}} f)_l(k)=\sum_{x_1=0}^{s_1-1} \cdots \sum_{x_d=0}^{s_d-1} f_l\left(x_1, \ldots, x_d\right) e^{-2 i \pi \sum_{j=1}^d \frac{x_j k_j}{s_j}}, \\
& \left(\hat{\mathcal{F}}^{-1} f\right)_l(x)=\sum_{k_1=0}^{s_1-1} \cdots \sum_{k_d=0}^{s_d-1} f_l\left(k_1, \ldots, k_d\right) e^{2 i \pi \sum_{j=1}^d \frac{x_j k_j}{s_j}}.
\end{align}

Finally, since Eq.~\eqref{eq:iter} follows standard neural network structures training a network training is done with an appropriate loss function  $\mathcal{L}=\mathcal{U} \times \mathcal{U}$
\begin{equation}
    \Theta=\underset{\Theta}{\arg \min } (\mathcal{L}(\mathcal{G}(a), G(a, \Theta)) .
\end{equation}

A schematic representation of the Fourier Neural Operator model for the Marshak wave problem is provided in Figure \ref{fig:FNO}.

\section{Results}\label{sec5}

\subsection{Problem description and parameter space}
The Marshak waves we consider concern the propagation of heat waves  through low-density foam cylinders or other materials driven by a hohlraum similar to those described in \cite{cohen2020key,fryer2016uncertainties}. Key parameters in these experiments include density, drive energy and radiation temperature, which typically can range from 100 to 300 eV. X-ray imaging is used to track the heat wave, while diagnostic tools measure the flux breaking through the foam edge. The experiments cover a wide range of temperatures, materials, and densities. 

Table \ref{tab:matprop}, adapted from \cite{cohen2020key}, presents material properties used in various Marshak wave experiments. The first ten rows contain parameters for the foams, while the last two rows provide parameters for coating materials.  
For each material, the numerical parameters were fitted in relevant experimental regimes. Further details about the experiments can be found in \cite{cohen2020key} and references cited therein.  
\begin{table}[htb!]
\centering
\caption{Material properties for various Marshak wave experiments }
\label{tab:matprop}
\begin{tabular}{llcclllll}
\toprule
Experiment & Foam & $g\left(\mathrm{g}/\mathrm{cm}^2\right)$ & $f\left(\mathrm{MJ}\right)$ & $\alpha$ & $\beta$ & $\lambda$ & $\mu$ & $\rho\left(\mathrm{g}/\mathrm{cm}^3\right)$ \\
\midrule
Massen & $\mathrm{C}_{11}\mathrm{H}_{16}\mathrm{Pb}_{0.3852}$ & $1/3200$ & 10.17 & 1.57 & 1.2 & 0.1 & 0 & 0.080 \\
Xu pure & $\mathrm{C}_6\mathrm{H}_{12}$ & $1/3926.6$ & 12.27 & 2.98 & 1 & 0.95 & 0.04 & 0.05 \\
Xu with copper & $\mathrm{C}_6\mathrm{H}_{12}\mathrm{Cu}_{0.394}$ & $1/7692.9$ & 8.13 & 3.44 & 1.1 & 0.67 & 0.07 & 0.05 \\
Back, Moore & $\mathrm{SiO}_2$ & $1/9175$ & 8.77 & 3.53 & 1.1 & 0.75 & 0.09 & 0.05 \\
Back & $\mathrm{Ta}_2\mathrm{O}_5$ & $1/8433.3$ & 4.78 & 1.78 & 1.37 & 0.24 & 0.12 & 0.04 \\
Back low energy & $\mathrm{SiO}_2$ & $1/9652$ & 8.4 & 2.0 & 1.23 & 0.61 & 0.1 & 0.01 \\
Moore & $\mathrm{C}_8\mathrm{H}_7\mathrm{Cl}$ & $1/24466$ & 14.47 & 5.7 & 0.96 & 0.72 & 0.04 & 0.105 \\
Keiter Pure & $\mathrm{C}_{15}\mathrm{H}_{20}\mathrm{O}_6$ & $1/26549$ & 11.54 & 5.29 & 0.94 & 0.95 & 0.038 & 0.065 \\
Keiter with Gold & $\mathrm{C}_{15}\mathrm{H}_{20}\mathrm{O}_6\mathrm{Au}_{0.172}$ & $1/4760$ & 9.81 & 2.5 & 1.04 & 0.35 & 0.06 & 0.0625 \\
Ji-Yan & $\mathrm{C}_8\mathrm{H}_8$ & $1/2818.1$ & 21.17 & 2.79 & 1.06 & 0.81 & 0.06 & 0.160 \\
& $\mathrm{Au}$ & $1/7200$ & 3.4 & 1.5 & 1.6 & 0.2 & 0.14 & 0.160 \\
& $\mathrm{Be}$ & $1/402.8$ & 8.81 & 4.89 & 1.09 & 0.67 & 0.07 & 0.160 \\
\bottomrule
\end{tabular}
\end{table}

Numerical approximations for solving the Marshak wave problem can be computationally expensive, especially when exploring a wide range of material properties. To overcome this challenge, we propose using the Fourier Neural Operator (FNO) to learn the mapping between material properties and their corresponding Marshak wave solutions.  FNOs have shown success in solving partial differential equations by learning the solution operator from a dataset of input-output pairs.  

To train the FNO model, we generate a dataset that spans the parameter space defined by the material properties in Table \ref{tab:matprop}. The input consists of a set of material properties, $(g, f, \alpha, \beta, \lambda, \mu, \rho)$, while the output corresponds to the solution of the Marshak wave problem in terms of the temperature profile and wave front position at a given time.  We create a uniformly spaced grid of values for each material property, covering the range of values found in the experiments: 
\begin{table}[htb!]
\centering
\caption{Parameter ranges for generating training data}
\label{tab:param_ranges}
\begin{tabular}{ccc}
\toprule
Parameter & Range & Number of grid points \\
\midrule
$g$ & $[\min(g), \max(g)]$ & $N$ (log-spaced) \\
$f$ & $[\min(f), \max(f)]$ & $N$ \\
$\alpha$ & $[\min(\alpha), \max(\alpha)]$ & $N$ \\
$\beta$ & $[\min(\beta), \max(\beta)]$ & $N$ \\
$\lambda$ & $[\min(\lambda), \max(\lambda)]$ & $N$ \\
$\mu$ & $[\min(\mu), \max(\mu)]$ & $N$ \\
$\rho$ & $[\min(\rho), \max(\rho)]$ & $N$ \\
\bottomrule
\end{tabular}
\end{table}
In Table \ref{tab:param_ranges}, $N$ is the number of grid points for each parameter. For the $g$ parameter, we use logarithmically spaced values to better capture its wide range, while the other parameters are linearly spaced. 

In addition to the material properties, the Marshak wave problem also depends on the boundary temperature (i.e., the drive temperature). We parameterize the drive with a function $T_b(t,a,b,c,d)$, measured in HeV, defined as follows
\begin{equation}\label{Tb}
    T_b(t,a,b,c,d) =  a + (b(t \geq c)(t-c))(t<d) + (t\geq d)(b(d - c)).
\end{equation}
Here $t$ is time (in ns), and $a \in [1, 3]$, $b \in [0, 1]$, $c \in [0.1, 2]$, and $d \in [2, 5]$. The function consists of a constant term $a$, and a piecewise function that takes different values based on the conditions involving $t$, $c$, and $d$. We generate a set of boundary temperature functions by sampling the parameters $a$, $b$, $c$, and $d$ from their respective ranges. 

To create the training set, we take the Cartesian product of the material property values and the boundary temperature function parameters and obtain a set of input parameter combinations that cover the entire parameter space.  For each input combination, we solve the Marshak wave problem using a numerical solver to obtain the corresponding output solution. These input-output pairs form our training dataset, which we use to train the FNO model.  

As will be seen, by learning from this diverse set of input-output pairs, the FNO can effectively capture the underlying physics of the Marshak wave problem across the entire parameter space, including the dependence on the boundary temperature function.  This allows the trained model to quickly and accurately predict solutions for new, unseen combinations of material properties and boundary temperature functions within the specified ranges.

\subsection{Base model}

As a starting point, we introduce a base model that takes all material properties and boundary temperature function parameters as inputs and uses the Hammer and Rosen approximation as the output. The Hammer and Rosen approximation provides an analytical solution to the Marshak wave problem, which serves as a useful benchmark for evaluating the performance of our FNO model. 

\begin{figure}[hbt!]
     \centering
     \begin{subfigure}[hbt!]{0.45\textwidth}
         \centering
         \includegraphics[width=\textwidth]{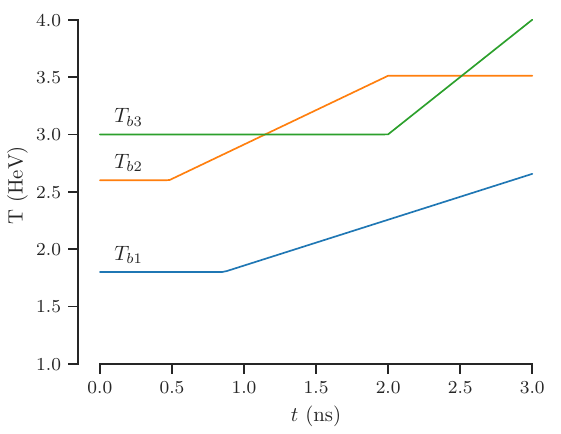}
         \caption{Temperature Drive}
         \label{fig:bt}
     \end{subfigure}
     \begin{subfigure}[hbt!]{0.45\textwidth}
         \centering
         \includegraphics[width=\textwidth]{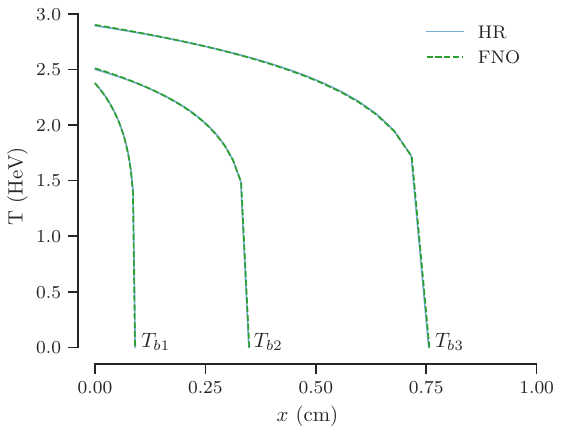}
         \caption{Temperature profile at 3 ns}
         \label{fig:solhrnn}
     \end{subfigure}
     \caption{Comparison of the Hammer and Rosen approximation and the FNO model for a representative material under different boundary temperature drives (a) are characterized by a constant temperature followed by a linear ramp at different times and rates.  The corresponding temperature solutions (b) obtained from the Hammer and Rosen approximation (solid lines) and the FNO model (dashed lines) show close agreement.}
     \label{fig:varybt}
\end{figure}

Figure \ref{fig:varybt} compares the temperature solutions of the Marshak wave in space for three different boundary temperature functions. The boundary temperature functions, shown in Figure \ref{fig:bt}, are generated by varying the parameters $a$, $b$, $c$, and $d$ in Equation \ref{Tb}. The corresponding temperature solutions, obtained using both the Hammer and Rosen approximation and the FNO model, are presented in Figure \ref{fig:solhrnn}.

The results demonstrate good agreement between the FNO model and the Hammer and Rosen approximation for all three boundary temperature functions. This indicates that the FNO model is capable of accurately capturing the physics of the Marshak wave problem and reproducing the analytical solutions provided by the Hammer and Rosen approximation.

\subsection{Hammer and Rosen Correction model}

While the Hammer and Rosen approximation provides an analytical solution to the Marshak wave problem, it suffers from inaccuracies due to the assumptions made in its derivation, Section \ref{sec3}. These inaccuracies become apparent when comparing the Hammer and Rosen solution to more accurate numerical solvers, such as diffusion based methods, and experimental results. To address this issue, we introduce the Hammer and Rosen Correction model, which aims to improve the accuracy of the Hammer and Rosen approximation using FNO. 

The Hammer and Rosen Correction model is built similarly to the base model but takes the Hammer and Rosen solution for the temperature and the front position as additional inputs. The outputs are generated using a more accurate diffusion solution, and the FNO learns to map the Hammer and Rosen solution to the diffusion solution. By doing so, the Hammer and Rosen Correction model effectively corrects the inaccuracies of the Hammer and Rosen approximation and provides a more accurate prediction of the Marshak wave behavior.

\begin{figure}[!htbp]
    \centering
    \includegraphics[width=0.45\textwidth]{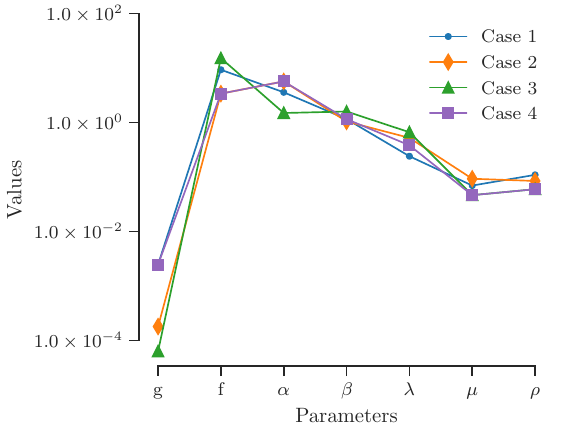}
    \caption{Parameter values from the test set for four different cases to evaluate the performance of the Hammer and Rosen Correction model}
    \label{fig:parvalues}
\end{figure}

Figure \ref{fig:parvalues} illustrates in a parallel axis plot the input parameter values for four different test cases used to evaluate the Hammer and Rosen Correction model. Each line represents a specific test case, with the values of the parameters plotted along the y-axis for each parameter on the x-axis. The boundary temperature drive is given with parameters $a = 1.2$, $b = 0.8$, $c = 1$, and $d = 2$ for Eq.~\eqref{Tb}.

The output values are produced by a numerical solver we developed to solve radiation diffusion in planar geometry. The solver assumes equilibrium between the radiation temperature and material temperature, reducing Eq.~\eqref{eq:rad} and Eq.~\eqref{eq:mat} to a single equation for the material temperature Eq.~\eqref{eq:diffusion_equilibrium}. The solver employs finite difference method to discretize the spatial domain into a uniform grid.  Time integration is performed by the backward differentiation formula, an implicit multi-step method.  The spatial derivatives in Eq.~\eqref{eq:diffusion_equilibrium} are approximated using a second order central difference scheme.  The left boundary at the surface ($x = 0$), Eq.~\eqref{eq:BC}, is prescribed as a function of time and the solver assumes equation of state given by Eq.~\eqref{eq:eos}. At each time step, the solver computes the temperature profile across a one-dimensional spatial grid consisting of 100 spatial cells and tracks the position of the wavefront.  

The Hammer and Rosen correction model is trained and tested using the dataset generated by the numerical solver and the Hammer and solution, paired with the input parameter values. The dataset is split into standard training and testing sets. It is important to note that the testing set contains parameter combinations that may not represent physically realistic scenarios, as they are generated by uniformly sampling the parameter space defined in Table \ref{tab:param_ranges}. The model is trained using 1.05M input-output pairs, with 58k trainable parameters and is trained over 30 epochs.

\begin{figure}[hbt!]
     \centering
     \begin{subfigure}[hbt!]{0.45\textwidth}
         \centering
         \includegraphics[width=\textwidth]{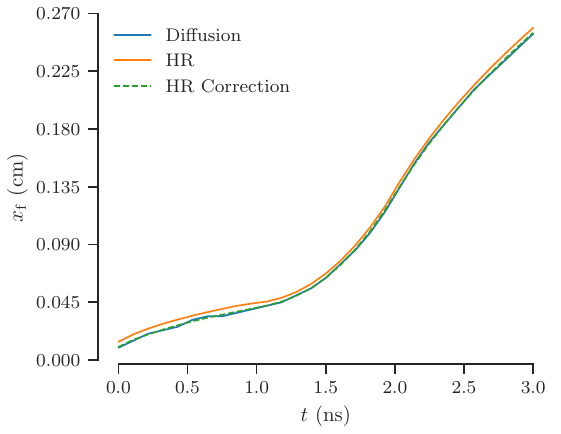}
         \caption{Case 1 front position solution}
         \label{fig:fp1}
     \end{subfigure}
     \begin{subfigure}[hbt!]{0.45\textwidth}
         \centering
         \includegraphics[width=\textwidth]{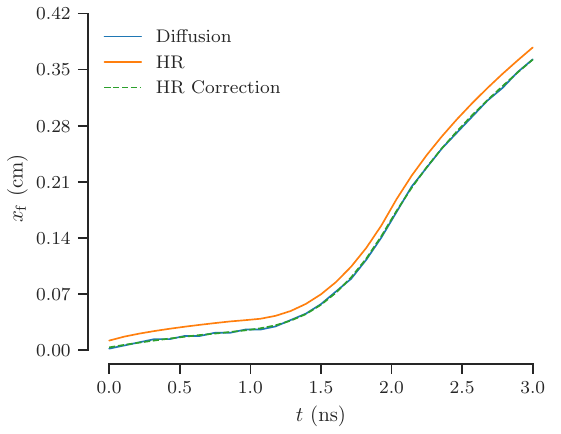}
         \caption{Case 2 front position solution}
         \label{fig:fp2}
     \end{subfigure}
     \begin{subfigure}[hbt!]{0.45\textwidth}
         \centering
         \includegraphics[width=\textwidth]{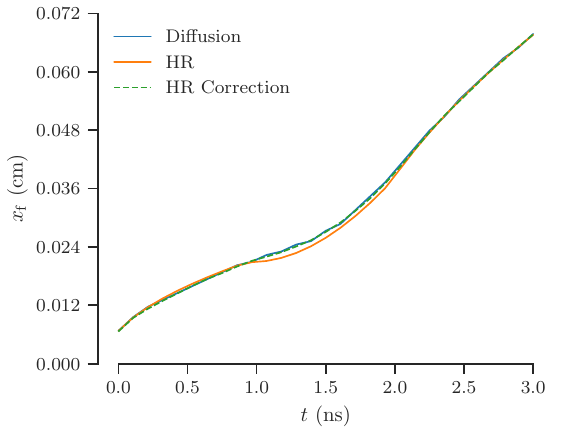}
         \caption{Case 3 front position solution}
         \label{fig:fp3}
     \end{subfigure}
     \begin{subfigure}[hbt!]{0.45\textwidth}
         \centering
         \includegraphics[width=\textwidth]{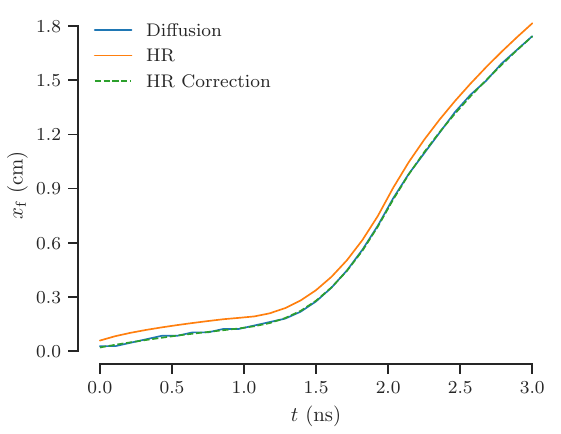}
         \caption{Case 4 front position solution}
         \label{fig:fp4}
     \end{subfigure}
     \caption{Comparison of the front position solutions over time for the Hammer and Rosen approximation, the Hammer and Rosen Correction model, and the diffusion solution for different sets of input parameters.  The Hammer and Rosen approximation (orange lines), deviates from the diffusion solution (blue lines) over time, while the Hammer and Rosen Correction (dashed green lines) accurately predicts the diffusion solution.}
     \label{fig:fp}
\end{figure}

Figure \ref{fig:fp} presents a comparison of the front position solutions over time for the Hammer and Rosen approximation, the Hammer and Rosen Correction model, and the diffusion solution.  The subfigures \ref{fig:fp1}, \ref{fig:fp2}, \ref{fig:fp3}, and \ref{fig:fp4} show the results for different sets of input parameters.  It is evident from the figures that the Hammer and Rosen approximation deviates noticeable from the diffusion solution over time. In contrast, the Hammer and Rosen Correction model accurately predicts the diffusion solution, demonstrating its ability to correct the inaccuracies of the Hammer and Rosen approximation. 

\begin{figure}[hbt!]
     \centering
     \begin{subfigure}[hbt!]{0.45\textwidth}
         \centering
         \includegraphics[width=\textwidth]{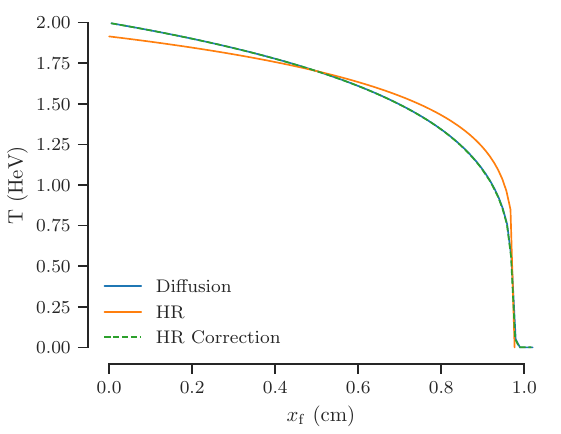}
         \caption{Case 1 temperature solution}
         \label{fig:t1}
     \end{subfigure}
     \begin{subfigure}[hbt!]{0.45\textwidth}
         \centering
         \includegraphics[width=\textwidth]{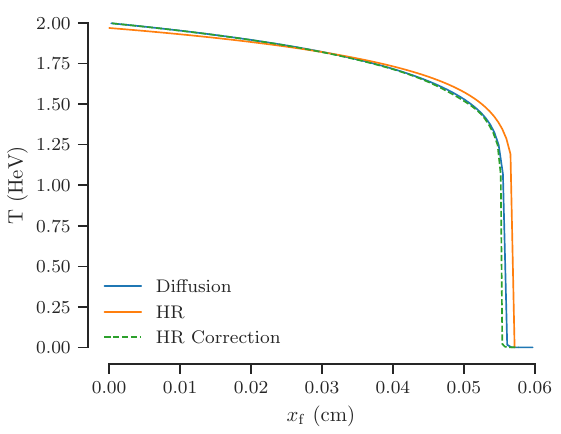}
         \caption{Case 2 temperature solution}
         \label{fig:t2}
     \end{subfigure}
     \begin{subfigure}[hbt!]{0.45\textwidth}
         \centering
         \includegraphics[width=\textwidth]{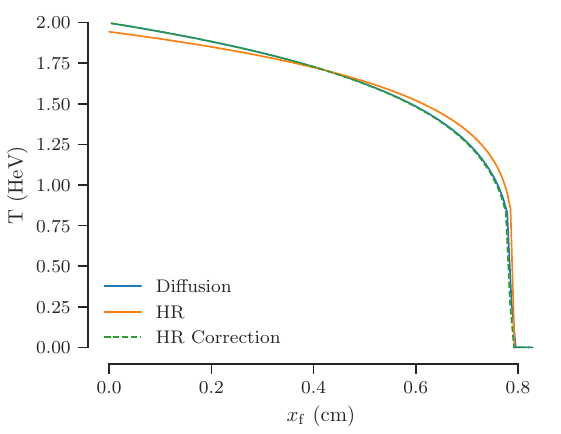}
         \caption{Case 3 temperature solution}
         \label{fig:t3}
     \end{subfigure}
     \begin{subfigure}[hbt!]{0.45\textwidth}
         \centering
         \includegraphics[width=\textwidth]{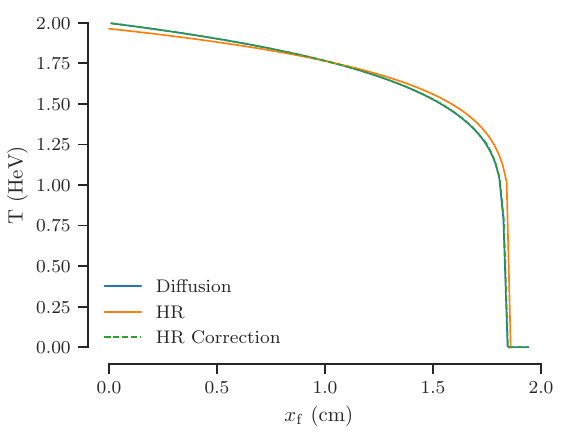}
         \caption{Case 4 temperature solution}
         \label{fig:t4}
     \end{subfigure}
     \caption{Comparison of the temperature profiles for the Hammer and Rosen approximation, the Hammer and Rosen Correction model, and the diffusion solution at the same time instance for different sets of input parameters.  The Hammer and Rosen approximation (orange line) exhibits discrepancies compared to the diffusion solution (blue line), while the Hammer and Rosen Correction (dashed green lines) closely match the diffusion solution.}
     \label{fig:t}
\end{figure}

Figure \ref{fig:t} provides a comparison of the temperature solutions for the same three models. Subfigures \ref{fig:t1}, \ref{fig:t2}, \ref{fig:t3}, and \ref{fig:t4} show the temperature profiles at the same time instance. Once again, the Hammer and Rosen Correction model closely matches the diffusion solution, while the Hammer and Rosen approximation exhibits discrepancies. 

The Hammer and Rosen Correction model  both improves the accuracy of the Marshak wave Hammer and Rosen solution and provides a framework for integrating analytical approximations with data-driven approaches. This hybrid approach combines benefits of both analytical and machine learning methods by giving a physical solution to simplify the inference.


\subsection{Model generalization and performance}

In the previous sections, we demonstrated the effectiveness of the Hammer and Rosen Correction model in accurately predicting the Marshak wave behavior for unseen data. It is important to note that these tests were performed on collocation points of the spacing grid shown in Table \ref{tab:param_ranges}. To validate generalization capabilities of FNO, we present additional tests on specific physical materials from Table \ref{tab:matprop}. 

\begin{figure}[hbt!]
     \centering
     \begin{subfigure}[hbt!]{0.45\textwidth}
         \centering
         \includegraphics[width=\textwidth]{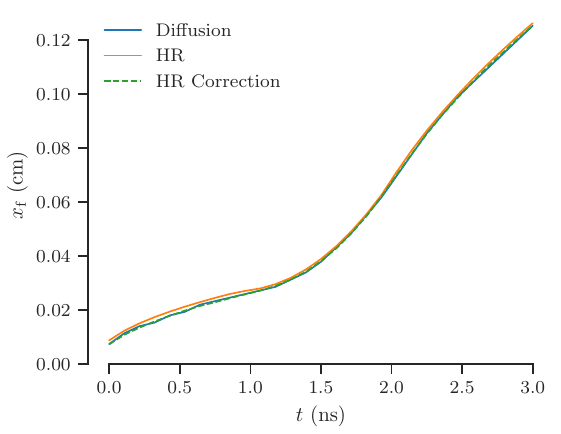}
         \caption{$\mathrm{C}_{15}\mathrm{H}_{20}\mathrm{O}_6\mathrm{Au}_{0.172}$}
         \label{fig:cm1}
     \end{subfigure}
     \begin{subfigure}[hbt!]{0.45\textwidth}
         \centering
         \includegraphics[width=\textwidth]{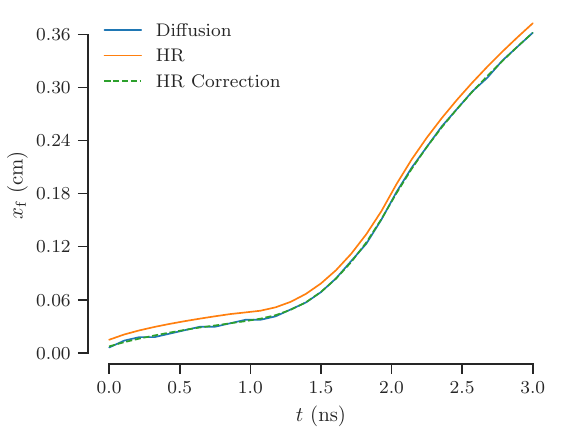}
         \caption{$\mathrm{Be}$}
         \label{fig:cm2}
     \end{subfigure}
     \begin{subfigure}[hbt!]{0.45\textwidth}
         \centering
         \includegraphics[width=\textwidth]{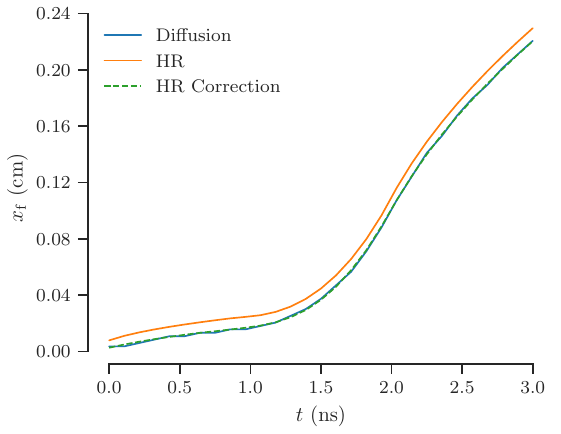}
         \caption{$\mathrm{C}_{15}\mathrm{H}_{20}\mathrm{O}_6$}
         \label{fig:cm3}
     \end{subfigure}
     \begin{subfigure}[hbt!]{0.45\textwidth}
         \centering
         \includegraphics[width=\textwidth]{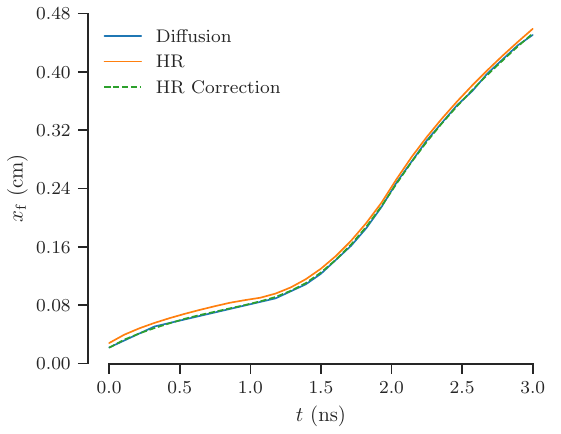}
         \caption{ $\mathrm{C}_6\mathrm{H}_{12}$}
         \label{fig:cm4}
     \end{subfigure}
     \caption{Comparison of the front positions obtained from the Hammer and Rosen approximation (orange lines), diffusion solver (blue lines), and the Hammer and Rosen Correction model (dashed green lines) for four different materials from the Table \ref{tab:matprop}.}
     \label{fig:cm}
\end{figure}

Figure \ref{fig:cm} compares the front position solutions obtained from the diffusion solver and the Hammer and Rosen Correction model for four different materials: $\mathrm{C}_{15}\mathrm{H}_{20}\mathrm{O}_6\mathrm{Au}_{0.172}$, $\mathrm{Be}$, $\mathrm{C}_{15}\mathrm{H}_{20}\mathrm{O}_6$, and $\mathrm{C}_6\mathrm{H}_{12}$ with properties as specified in \cite{cohen2020key}. These materials were not explicitly included in the training data grid but represent realistic physical scenarios.  The subfigures \ref{fig:cm1}, \ref{fig:cm2}, \ref{fig:cm3}, and \ref{fig:cm4} show excellent agreement between diffusion solutions and the Hammer and Rosen Correction model predictions for all four materials. This demonstrates that the FNO has successfully learned the mapping in the entire parameter space and can accurately predict the Marshak wave behavior for arbitrary material properties within the considered ranges. 

To quantitatively asses the performance and computational efficiency of the Hammer and Rosen Correction model, we compare it with the base model in Table \ref{tab:comparison}. Both models are trained with the same number of trainable parameters, training data, and epochs to ensure a fair comparison.  The mean squared error (MSE) is used as the evaluation metric for both temperature and front position predictions.  

\begin{table}[htb!]
\centering
\caption{Prediction performance and computational costs of deep learning models (MSE is the mean squared error)}
\label{tab:comparison}
\begin{tabular}{l@{\hspace{1cm}}lll}
\toprule
Parameter & HR Correction & Base model & \% Improvement \\
\midrule
Temperature MSE & \textbf{0.00081} & 0.00185 & 56.16 \\
\addlinespace
Front position MSE & \textbf{0.00807} & 0.01220 & 33.93 \\
\addlinespace
Train data  & 1.05M & 1.05M &  \\
Trainable parameters & 58k & 58k &  \\
Epochs & 30 & 30 &  \\
Inference time (s) & 0.0032 & \textbf{0.0016} &  \\
\bottomrule
\end{tabular}
\end{table}

The results in Table \ref{tab:comparison} show that the Hammer and Rosen Correction model significantly outperforms the base model in terms of prediction accuracy.  The Hammer and Rosen Correction model achieves a 56.16\% improvement in temperature MSE and a 33.93\% improvement in front position MSE compared to the base model. This superior performance can be attributed to the hybrid-nature approach of the Hammer and Rosen Correction model.  

In terms of computational efficiency, the Hammer and Rosen Correction model has slightly slower inference time as compared to the base model.  This is expected due to the additional complexity introduced by the correction step. However, it is important to note that both models have extremely fast inference times, with the Hammer and Rosen Correction model requiring only 0.0032 seconds per prediction and the base model requiring 0.0016 seconds. These fast inference time highlight the efficiency of the FNO-based approach, enabling real-time predictions of the Marshak wave behavior.

\section{Conclusion}\label{sec6}

In this work, we presented a novel approach for modeling Marshak wave experiments using Fourier Neural Operators (FNO). The primary objective was to develop an efficient and accurate method for predicting Marshak wave behavior across a wide range of material properties and boundary temperature functions. We introduced two FNO-based models: a base model and a Hammer and Rosen Correction model. The base model takes material properties and boundary temperature function parameters as inputs and uses a numerical approximation as the output. This model served as a foundation for exploring the capabilities of learning the underlying physics. To address innaccuracies of the Hammer and Rosen approximation, we developed a hybrid data-driven Hammer and Rosen Correction model. This model maps the Hammer and Rosen solution to a more accurate diffusion solution. The performance of these models were evaluated over a wide range of the parameter space. The results demonstrated strong generalization capabilities on unseen data. The Hammer and Rosen Correction model achieved 56.16\% improvement in temperature MSE and a 33.93\% improvement in front position MSE compared to the base model. These results pave the way for further exploration of more complex models and application to multidimensional problems in high energy density physics.

\bibliography{sn-article}


\begin{thebibliography}{32}
\ifx \bisbn   \undefined \def \bisbn  #1{ISBN #1}\fi
\ifx \binits  \undefined \def \binits#1{#1}\fi
\ifx \bauthor  \undefined \def \bauthor#1{#1}\fi
\ifx \batitle  \undefined \def \batitle#1{#1}\fi
\ifx \bjtitle  \undefined \def \bjtitle#1{#1}\fi
\ifx \bvolume  \undefined \def \bvolume#1{\textbf{#1}}\fi
\ifx \byear  \undefined \def \byear#1{#1}\fi
\ifx \bissue  \undefined \def \bissue#1{#1}\fi
\ifx \bfpage  \undefined \def \bfpage#1{#1}\fi
\ifx \blpage  \undefined \def \blpage #1{#1}\fi
\ifx \burl  \undefined \def \burl#1{\textsf{#1}}\fi
\ifx \doiurl  \undefined \def \doiurl#1{\url{https://doi.org/#1}}\fi
\ifx \betal  \undefined \def \betal{\textit{et al.}}\fi
\ifx \binstitute  \undefined \def \binstitute#1{#1}\fi
\ifx \binstitutionaled  \undefined \def \binstitutionaled#1{#1}\fi
\ifx \bctitle  \undefined \def \bctitle#1{#1}\fi
\ifx \beditor  \undefined \def \beditor#1{#1}\fi
\ifx \bpublisher  \undefined \def \bpublisher#1{#1}\fi
\ifx \bbtitle  \undefined \def \bbtitle#1{#1}\fi
\ifx \bedition  \undefined \def \bedition#1{#1}\fi
\ifx \bseriesno  \undefined \def \bseriesno#1{#1}\fi
\ifx \blocation  \undefined \def \blocation#1{#1}\fi
\ifx \bsertitle  \undefined \def \bsertitle#1{#1}\fi
\ifx \bsnm \undefined \def \bsnm#1{#1}\fi
\ifx \bsuffix \undefined \def \bsuffix#1{#1}\fi
\ifx \bparticle \undefined \def \bparticle#1{#1}\fi
\ifx \barticle \undefined \def \barticle#1{#1}\fi
\bibcommenthead
\ifx \bconfdate \undefined \def \bconfdate #1{#1}\fi
\ifx \botherref \undefined \def \botherref #1{#1}\fi
\ifx \url \undefined \def \url#1{\textsf{#1}}\fi
\ifx \bchapter \undefined \def \bchapter#1{#1}\fi
\ifx \bbook \undefined \def \bbook#1{#1}\fi
\ifx \bcomment \undefined \def \bcomment#1{#1}\fi
\ifx \oauthor \undefined \def \oauthor#1{#1}\fi
\ifx \citeauthoryear \undefined \def \citeauthoryear#1{#1}\fi
\ifx \endbibitem  \undefined \def \endbibitem {}\fi
\ifx \bconflocation  \undefined \def \bconflocation#1{#1}\fi
\ifx \arxivurl  \undefined \def \arxivurl#1{\textsf{#1}}\fi
\csname PreBibitemsHook\endcsname

\bibitem[\protect\citeauthoryear{Lindl et~al.}{2004}]{lindl2004physics}
\begin{barticle}
\bauthor{\bsnm{Lindl}, \binits{J.D.}},
\bauthor{\bsnm{Amendt}, \binits{P.}},
\bauthor{\bsnm{Berger}, \binits{R.L.}},
\bauthor{\bsnm{Glendinning}, \binits{S.G.}},
\bauthor{\bsnm{Glenzer}, \binits{S.H.}},
\bauthor{\bsnm{Haan}, \binits{S.W.}},
\bauthor{\bsnm{Kauffman}, \binits{R.L.}},
\bauthor{\bsnm{Landen}, \binits{O.L.}},
\bauthor{\bsnm{Suter}, \binits{L.J.}}:
\batitle{The physics basis for ignition using indirect-drive targets on the national ignition facility}.
\bjtitle{Physics of plasmas}
\bvolume{11}(\bissue{2}),
\bfpage{339}--\blpage{491}
(\byear{2004})
\end{barticle}
\endbibitem

\bibitem[\protect\citeauthoryear{Rosen and Hammer}{2005}]{rosen2005analytic}
\begin{barticle}
\bauthor{\bsnm{Rosen}, \binits{M.D.}},
\bauthor{\bsnm{Hammer}, \binits{J.H.}}:
\batitle{Analytic expressions for optimal inertial-confinement-fusion hohlraum wall density and wall loss}.
\bjtitle{Physical Review E}
\bvolume{72}(\bissue{5}),
\bfpage{056403}
(\byear{2005})
\end{barticle}
\endbibitem

\bibitem[\protect\citeauthoryear{Cohen et~al.}{2004}]{cohen2004tracer}
\begin{barticle}
\bauthor{\bsnm{Cohen}, \binits{D.H.}},
\bauthor{\bsnm{MacFarlane}, \binits{J.J.}},
\bauthor{\bsnm{Jaanimagi}, \binits{P.}},
\bauthor{\bsnm{Landen}, \binits{O.L.}},
\bauthor{\bsnm{Haynes}, \binits{D.A.}},
\bauthor{\bsnm{Conners}, \binits{D.S.}},
\bauthor{\bsnm{Penrose}, \binits{K.L.}},
\bauthor{\bsnm{Shupe}, \binits{N.C.}}:
\batitle{Tracer spectroscopy diagnostics of doped ablators in inertial confinement fusion experiments on omega}.
\bjtitle{Physics of Plasmas}
\bvolume{11}(\bissue{5}),
\bfpage{2702}--\blpage{2708}
(\byear{2004})
\end{barticle}
\endbibitem

\bibitem[\protect\citeauthoryear{Lindl}{1995}]{lindl1995development}
\begin{barticle}
\bauthor{\bsnm{Lindl}, \binits{J.}}:
\batitle{Development of the indirect-drive approach to inertial confinement fusion and the target physics basis for ignition and gain}.
\bjtitle{Physics of plasmas}
\bvolume{2}(\bissue{11}),
\bfpage{3933}--\blpage{4024}
(\byear{1995})
\end{barticle}
\endbibitem

\bibitem[\protect\citeauthoryear{Tranchant et~al.}{2022}]{tranchant2022new}
\begin{barticle}
\bauthor{\bsnm{Tranchant}, \binits{V.}},
\bauthor{\bsnm{Charpentier}, \binits{N.}},
\bauthor{\bsnm{Som}, \binits{L.V.B.}},
\bauthor{\bsnm{Ciardi}, \binits{A.}},
\bauthor{\bsnm{Falize}, \binits{{\'E}.}}:
\batitle{New class of laboratory astrophysics experiments: Application to radiative accretion processes around neutron stars}.
\bjtitle{The Astrophysical Journal}
\bvolume{936}(\bissue{1}),
\bfpage{14}
(\byear{2022})
\end{barticle}
\endbibitem

\bibitem[\protect\citeauthoryear{Gonz{\'a}lez et~al.}{2009}]{gonzalez20092d}
\begin{barticle}
\bauthor{\bsnm{Gonz{\'a}lez}, \binits{M.}},
\bauthor{\bsnm{Audit}, \binits{E.}},
\bauthor{\bsnm{Stehl{\'e}}, \binits{C.}}:
\batitle{2d numerical study of the radiation influence on shock structure relevant to laboratory astrophysics}.
\bjtitle{Astronomy \& Astrophysics}
\bvolume{497}(\bissue{1}),
\bfpage{27}--\blpage{34}
(\byear{2009})
\end{barticle}
\endbibitem

\bibitem[\protect\citeauthoryear{Rosen}{1996}]{rosen1996science}
\begin{barticle}
\bauthor{\bsnm{Rosen}, \binits{M.D.}}:
\batitle{The science applications of the high-energy density plasmas created on the nova laser}.
\bjtitle{Physics of Plasmas}
\bvolume{3}(\bissue{5}),
\bfpage{1803}--\blpage{1812}
(\byear{1996})
\end{barticle}
\endbibitem

\bibitem[\protect\citeauthoryear{Drake}{2006}]{Drake}
\begin{bbook}
\bauthor{\bsnm{Drake}, \binits{R.P.}}:
\bbtitle{{High Energy Density Physics}}.
\bsertitle{Springer}.
\bpublisher{Springer}, \blocation{???}
(\byear{2006})
\end{bbook}
\endbibitem

\bibitem[\protect\citeauthoryear{Olson et~al.}{2000}]{olson2000diffusion}
\begin{barticle}
\bauthor{\bsnm{Olson}, \binits{G.L.}},
\bauthor{\bsnm{Auer}, \binits{L.H.}},
\bauthor{\bsnm{Hall}, \binits{M.L.}}:
\batitle{Diffusion, p1, and other approximate forms of radiation transport}.
\bjtitle{Journal of Quantitative Spectroscopy and Radiative Transfer}
\bvolume{64}(\bissue{6}),
\bfpage{619}--\blpage{634}
(\byear{2000})
\end{barticle}
\endbibitem

\bibitem[\protect\citeauthoryear{Hurricane and Hammer}{2006}]{hurricane2006bent}
\begin{botherref}
\oauthor{\bsnm{Hurricane}, \binits{O.}},
\oauthor{\bsnm{Hammer}, \binits{J.}}:
Bent marshak waves.
Physics of plasmas
\textbf{13}(11)
(2006)
\end{botherref}
\endbibitem

\bibitem[\protect\citeauthoryear{Hammer and Rosen}{2003}]{hammer2003consistent}
\begin{barticle}
\bauthor{\bsnm{Hammer}, \binits{J.H.}},
\bauthor{\bsnm{Rosen}, \binits{M.D.}}:
\batitle{A consistent approach to solving the radiation diffusion equation}.
\bjtitle{Physics of Plasmas}
\bvolume{10}(\bissue{5}),
\bfpage{1829}--\blpage{1845}
(\byear{2003})
\end{barticle}
\endbibitem

\bibitem[\protect\citeauthoryear{Heizler et~al.}{2016}]{heizler2016self}
\begin{barticle}
\bauthor{\bsnm{Heizler}, \binits{S.I.}},
\bauthor{\bsnm{Shussman}, \binits{T.}},
\bauthor{\bsnm{Malka}, \binits{E.}}:
\batitle{Self-similar solution of the subsonic radiative heat equations using a binary equation of state}.
\bjtitle{Journal of Computational and Theoretical Transport}
\bvolume{45}(\bissue{4}),
\bfpage{256}--\blpage{267}
(\byear{2016})
\end{barticle}
\endbibitem

\bibitem[\protect\citeauthoryear{Marshak}{1958}]{marshak}
\begin{botherref}
\oauthor{\bsnm{Marshak}, \binits{R.E.}}:
Effect of radiation on shock wave behavior.
The Physics of Fluids
\textbf{1}(1)
(1958)
\end{botherref}
\endbibitem

\bibitem[\protect\citeauthoryear{Petschek et~al.}{1960}]{petschek1960penetration}
\begin{botherref}
\oauthor{\bsnm{Petschek}, \binits{A.G.}},
\oauthor{\bsnm{Williamson}, \binits{R.E.}},
\oauthor{\bsnm{Wooten~Jr.}, \binits{J.K.}}:
The penetration of radiation with constant driving temperature.
Technical report,
Los Alamos Scientific Lab.
(1960)
\end{botherref}
\endbibitem

\bibitem[\protect\citeauthoryear{Lane and McClarren}{2013}]{lane2013new}
\begin{barticle}
\bauthor{\bsnm{Lane}, \binits{T.K.}},
\bauthor{\bsnm{McClarren}, \binits{R.G.}}:
\batitle{New self-similar radiation-hydrodynamics solutions in the high-energy density, equilibrium diffusion limit}.
\bjtitle{New Journal of Physics}
\bvolume{15}(\bissue{9}),
\bfpage{095013}
(\byear{2013})
\end{barticle}
\endbibitem

\bibitem[\protect\citeauthoryear{Krief and McClarren}{2024}]{krief2024self}
\begin{botherref}
\oauthor{\bsnm{Krief}, \binits{M.}},
\oauthor{\bsnm{McClarren}, \binits{R.G.}}:
Self-similar solutions for the non-equilibrium nonlinear supersonic marshak wave problem.
Physics of Fluids
\textbf{36}(1)
(2024)
\end{botherref}
\endbibitem

\bibitem[\protect\citeauthoryear{Renardy and Rogers}{2006}]{renardy2006introduction}
\begin{bbook}
\bauthor{\bsnm{Renardy}, \binits{M.}},
\bauthor{\bsnm{Rogers}, \binits{R.C.}}:
\bbtitle{An Introduction to Partial Differential Equations, Volume 13}.
\bpublisher{Springer}, \blocation{???}
(\byear{2006})
\end{bbook}
\endbibitem

\bibitem[\protect\citeauthoryear{Jones et~al.}{2009}]{jones2009differential}
\begin{bbook}
\bauthor{\bsnm{Jones}, \binits{D.S.}},
\bauthor{\bsnm{Plank}, \binits{M.}},
\bauthor{\bsnm{Sleeman}, \binits{B.D.}}:
\bbtitle{Differential Equations and Mathematical Biology}.
\bpublisher{Chapman and Hall/CRC}, \blocation{???}
(\byear{2009})
\end{bbook}
\endbibitem

\bibitem[\protect\citeauthoryear{Sommerfeld}{1949}]{sommerfeld1949partial}
\begin{bbook}
\bauthor{\bsnm{Sommerfeld}, \binits{A.}}:
\bbtitle{Partial Differential Equations in Physics}.
\bpublisher{Academic press}, \blocation{???}
(\byear{1949})
\end{bbook}
\endbibitem

\bibitem[\protect\citeauthoryear{Hughes}{2012}]{hughes2012finite}
\begin{bbook}
\bauthor{\bsnm{Hughes}, \binits{T.J.}}:
\bbtitle{The Finite Element Method: Linear Static and Dynamic Finite Element Analysis}.
\bpublisher{Courier Corporation}, \blocation{???}
(\byear{2012})
\end{bbook}
\endbibitem

\bibitem[\protect\citeauthoryear{Strikwerda}{2004}]{strikwerda2004finite}
\begin{bbook}
\bauthor{\bsnm{Strikwerda}, \binits{J.C.}}:
\bbtitle{Finite Difference Schemes and Partial Differential Equations}.
\bpublisher{SIAM}, \blocation{???}
(\byear{2004})
\end{bbook}
\endbibitem

\bibitem[\protect\citeauthoryear{Eymard et~al.}{2000}]{eymard2000finite}
\begin{bchapter}
\bauthor{\bsnm{Eymard}, \binits{R.}},
\bauthor{\bsnm{Gallouët}, \binits{T.}},
\bauthor{\bsnm{Herbin}, \binits{R.}}:
\bctitle{Finite volume methods}.
In: \bbtitle{Handbook of Numerical Analysis}
vol. \bseriesno{7},
pp. \bfpage{713}--\blpage{1018}.
\bpublisher{North-Holland}, \blocation{???}
(\byear{2000})
\end{bchapter}
\endbibitem

\bibitem[\protect\citeauthoryear{Wu and Xiu}{2020}]{wu2020data}
\begin{barticle}
\bauthor{\bsnm{Wu}, \binits{K.}},
\bauthor{\bsnm{Xiu}, \binits{D.}}:
\batitle{Data-driven deep learning of partial differential equations in modal space}.
\bjtitle{Journal of Computational Physics}
\bvolume{408},
\bfpage{109307}
(\byear{2020})
\end{barticle}
\endbibitem

\bibitem[\protect\citeauthoryear{Navaneeth and Chakraborty}{2022}]{navaneeth2022koopman}
\begin{botherref}
\oauthor{\bsnm{Navaneeth}, \binits{N.}},
\oauthor{\bsnm{Chakraborty}, \binits{S.}}:
Koopman operator for time-dependent reliability analysis.
arXiv e-prints,
2203
(2022)
\end{botherref}
\endbibitem

\bibitem[\protect\citeauthoryear{Goswami et~al.}{2020}]{goswami2020transfer}
\begin{barticle}
\bauthor{\bsnm{Goswami}, \binits{S.}},
\bauthor{\bsnm{Anitescu}, \binits{C.}},
\bauthor{\bsnm{Chakraborty}, \binits{S.}},
\bauthor{\bsnm{Rabczuk}, \binits{T.}}:
\batitle{Transfer learning enhanced physics informed neural network for phase-field modeling of fracture}.
\bjtitle{Theoretical and Applied Fracture Mechanics}
\bvolume{106},
\bfpage{102447}
(\byear{2020})
\end{barticle}
\endbibitem

\bibitem[\protect\citeauthoryear{Li et~al.}{2020}]{li2020neural}
\begin{botherref}
\oauthor{\bsnm{Li}, \binits{Z.}},
\oauthor{\bsnm{Kovachki}, \binits{N.}},
\oauthor{\bsnm{Azizzadenesheli}, \binits{K.}},
\oauthor{\bsnm{Liu}, \binits{B.}},
\oauthor{\bsnm{Bhattacharya}, \binits{K.}},
\oauthor{\bsnm{Stuart}, \binits{A.}},
\oauthor{\bsnm{Anandkumar}, \binits{A.}}:
Neural operator: Graph kernel network for partial differential equations
(2020)
\end{botherref}
\endbibitem

\bibitem[\protect\citeauthoryear{Lu et~al.}{2022}]{lu2022comprehensive}
\begin{barticle}
\bauthor{\bsnm{Lu}, \binits{L.}},
\bauthor{\bsnm{Meng}, \binits{X.}},
\bauthor{\bsnm{Cai}, \binits{S.}},
\bauthor{\bsnm{Mao}, \binits{Z.}},
\bauthor{\bsnm{Goswami}, \binits{S.}},
\bauthor{\bsnm{Zhang}, \binits{Z.}},
\bauthor{\bsnm{Karniadakis}, \binits{G.E.}}:
\batitle{A comprehensive and fair comparison of two neural operators (with practical extensions) based on fair data}.
\bjtitle{Computer Methods in Applied Mechanics and Engineering}
\bvolume{393},
\bfpage{114778}
(\byear{2022})
\end{barticle}
\endbibitem

\bibitem[\protect\citeauthoryear{Li et~al.}{2020}]{li2020fourier}
\begin{botherref}
\oauthor{\bsnm{Li}, \binits{Z.}},
\oauthor{\bsnm{Kovachki}, \binits{N.}},
\oauthor{\bsnm{Azizzadenesheli}, \binits{K.}},
\oauthor{\bsnm{Liu}, \binits{B.}},
\oauthor{\bsnm{Bhattacharya}, \binits{K.}},
\oauthor{\bsnm{Stuart}, \binits{A.}},
\oauthor{\bsnm{Anandkumar}, \binits{A.}}:
Fourier neural operator for parametric partial differential equations
(2020)
\end{botherref}
\endbibitem

\bibitem[\protect\citeauthoryear{Bingjing and Olson}{1996}]{bingjing1996benchmark}
\begin{barticle}
\bauthor{\bsnm{Bingjing}, \binits{S.}},
\bauthor{\bsnm{Olson}, \binits{G.L.}}:
\batitle{Benchmark results for the non-equilibrium marshak diffusion problem}.
\bjtitle{Journal of Quantitative Spectroscopy and Radiative Transfer}
\bvolume{56}(\bissue{3}),
\bfpage{337}--\blpage{351}
(\byear{1996})
\end{barticle}
\endbibitem

\bibitem[\protect\citeauthoryear{Cohen et~al.}{2020}]{cohen2020key}
\begin{barticle}
\bauthor{\bsnm{Cohen}, \binits{A.P.}},
\bauthor{\bsnm{Malamud}, \binits{G.}},
\bauthor{\bsnm{Heizler}, \binits{S.I.}}:
\batitle{Key to understanding supersonic radiative marshak waves using simple models and advanced simulations}.
\bjtitle{Physical Review Research}
\bvolume{2}(\bissue{2}),
\bfpage{023007}
(\byear{2020})
\end{barticle}
\endbibitem

\bibitem[\protect\citeauthoryear{Tripura and Chakraborty}{2022}]{tripura2022wavelet}
\begin{botherref}
\oauthor{\bsnm{Tripura}, \binits{T.}},
\oauthor{\bsnm{Chakraborty}, \binits{S.}}:
Wavelet neural operator: a neural operator for parametric partial differential equations
(2022)
\end{botherref}
\endbibitem

\bibitem[\protect\citeauthoryear{Fryer et~al.}{2016}]{fryer2016uncertainties}
\begin{barticle}
\bauthor{\bsnm{Fryer}, \binits{C.}},
\bauthor{\bsnm{Dodd}, \binits{E.}},
\bauthor{\bsnm{Even}, \binits{W.}},
\bauthor{\bsnm{Fontes}, \binits{C.}},
\bauthor{\bsnm{Greeff}, \binits{C.}},
\bauthor{\bsnm{Hungerford}, \binits{A.}},
\bauthor{\bsnm{Kline}, \binits{J.}},
\bauthor{\bsnm{Mussack}, \binits{K.}},
\bauthor{\bsnm{Tregillis}, \binits{I.}},
\bauthor{\bsnm{Workman}, \binits{J.}}, \betal:
\batitle{Uncertainties in radiation flow experiments}.
\bjtitle{High energy density physics}
\bvolume{18},
\bfpage{45}--\blpage{54}
(\byear{2016})
\end{barticle}
\endbibitem

\end{thebibliography}

\end{document}